\documentclass[prl,aps,twocolumn,preprintnumbers]{revtex4}

\newcommand{\checked}[1]{}
\newcommand{\comment}[1]{}

\usepackage{graphicx}
\usepackage{rotating}
\newcommand{\beq}{\begin{equation}}
\newcommand{\eeq}{\end{equation}}
\newcommand{\bqa}{\begin{eqnarray}}
\newcommand{\eqa}{\end{eqnarray}}

\def\sumint{\hbox{$\sum$}\!\!\!\!\!\!\int}

\begin{document}

\title{The Role of Quark Mass in Cold and Dense Perturbative QCD}
\preprint{BI-TP 2004/39}
\author{Eduardo S. {\sc Fraga}$^1$    
and Paul {\sc Romatschke}$^2$}
\affiliation{$^1$Instituto de F\'\i sica, 
Universidade Federal do Rio de Janeiro\\ 
C.P. 68528, Rio de Janeiro, RJ 21941-972, Brazil\\
$^2$Fakult\"at f\"ur Physik, Universit\"at Bielefeld, 
D-33615 Bielefeld, Germany}
\date{\today}

\begin{abstract}
We consider the equation of state of QCD at high density and zero 
temperature in perturbation theory to first order in the 
coupling constant $\alpha_s$. We compute the thermodynamic potential 
including the effect of a non-vanishing mass for the 
strange quark and show that corrections are sizable. Renormalization 
group running of the coupling and the strange quark mass plays a 
crucial role. The structure of quark stars is dramatically modified.
\end{abstract}
\maketitle

The investigation of the equation of state for cold and dense strongly 
interacting matter and its consequences for the possible phases of 
quantum chromodynamics (QCD) is still in its infancy. From the 
observational point of view, compact stars provide the most promising 
``laboratory'', since their central densities might be high enough to 
allow for the presence of deconfined quark matter. These dense objects 
are usually called quark stars or hybrid stars 
\cite{G_b}, and some observables, such as their mass-radius relation, 
may be used to constrain the equation of state for strong interactions.

In this Rapid Communication we compute the thermodynamic potential for 
cold quark matter with two light (massless) flavors, corresponding to the 
up and down quarks, and one massive flavor, corresponding to the 
strange quark, in perturbation theory to first order in $\alpha_s$ in the 
$\overline{\hbox{\scriptsize MS}}$ scheme. In this fashion, we can easily 
include modern renormalization group running effects 
for both the coupling and the strange quark mass. We find that 
the corrections due to the running nonzero mass are sizable, and should 
not be neglected in the evaluation of thermodynamic quantities. Solving the 
Tolman-Oppenheimer-Volkov (TOV) equations using our equation of state 
in the presence of electrons, 
we show that a running strange quark mass dramatically 
modifies the mass-radius diagram for quark stars, even at first order 
in $\alpha_s$. Although the numbers we present for the mass and  
radius of quark stars will be modified in a more realistic 
description, our results are consistent with constraints on the 
mass-to-radius ratio coming from measurements of gravitationally 
redshifted absorption lines in the X-ray burst spectra of the 
low-mass binary EXO$0748-676$ \cite{Cottam:2002cu}.

The thermodynamic potential for cold QCD in perturbation theory to 
$\sim\alpha_s^2$ was first computed a long time ago 
\cite{Free,Baluni,Toimela}. Nevertheless, 
the original approach to quark stars \cite{wit,bag} made use of 
the bag model with corrections $\sim\alpha_s$ from perturbative QCD to 
compute the thermodynamic potential 
\footnote{Actually, the authors of \cite{bag} considered the structure of 
strange stars in which, as suggested by Bodmer and Witten \cite{bod,wit}, 
strange quark matter would be the true ground state of QCD. In this work, we 
consider quark stars, so that a realistic description should match 
our equation 
for high density QCD onto a hadronic equation of state at lower densities.}. 
In the massless case, first-order corrections cancel out in the 
equation of state, 
so that one ends up with a free gas of quarks modified only 
by a bag constant.  
Finite quark mass effects were then estimated to modify the equation of 
state by less than $5\%$ and were essentially ignored. 

A few years ago, corrections $\sim\alpha_s^2$ 
with a modern definition of the running coupling constant were 
used to model the non-ideality in the equation of state for cold, dense 
QCD with massless quarks \cite{Fraga:2001id}. A perturbative calculation 
of the thermodynamic potential necessarily produces an unknown scale, 
$\bar{\Lambda}$, associated with the subtraction point for 
renormalization. In the case of zero temperature and mass 
and nonzero chemical 
potential, $\bar{\Lambda}$ is proportional to the quark chemical 
potential, $\mu$. Nevertheless, on very general grounds, one can argue that 
reasonable values for $\bar{\Lambda}/\mu$ lie between $2$ and $3$, if one 
takes perturbative QCD as a model for the equation of state 
\cite{Fraga:2001id} 
for cold strongly interacting matter. The former value corresponds 
to a strong 
first-order chiral phase transition, and the latter to a weak transition or a 
crossover. In the case of a strong first-order transition it was 
shown that there 
might be a new class of compact stars, besides the usual neutron stars, 
being smaller and much denser, with a large fraction 
of deconfined quark matter \cite{Fraga:2001id,Fraga:2001xc}.
This approach can be compared to treatments that resort to 
resummation methods and quasiparticle model descriptions 
\cite{andre,tony,andersen,paul}. Remarkably, these different frameworks seem 
to agree reasonably well for $\mu >> 1~$GeV, and point in 
the same direction even for $\mu \sim 1~$GeV and smaller, where one is 
clearly pushing perturbative QCD far below its region of applicability. 

Even the most recent QCD approaches mentioned above generally neglected 
quark masses and the presence of a color superconducting gap as compared 
to the typical scale for the chemical potential in the 
interior of compact stars, 
$\sim 400~$MeV and higher. However, it was recently argued that both effects 
should matter in the lower-density sector of the equation of 
state \cite{alford}. 
In fact, although quarks are essentially massless in the core of quark stars, 
the mass of the strange quark runs up, and becomes comparable to the typical 
scale for the chemical potential, as one approaches the surface of the star. 

In what follows, we present an exploratory analysis of the effects 
of a finite mass 
for the strange quark on the equation of state for perturbative QCD 
at high density, 
leaving the inclusion of a color superconducting gap in this framework 
for future investigations. To illustrate the effects and study the 
modifications in the 
structure of quark stars, we focus on the simpler case of first-order 
corrections. 
Results including full corrections $\sim\alpha_s^2$, as well as 
technical details 
of the calculation at each order and renormalization, will be 
presented in a longer 
publication \cite{next}.

The leading-order piece of the thermodynamic potential of QCD for one massive 
flavor is given by \cite{Free,Baluni,Toimela}

\beq
\Omega^{(0)}= - \frac{N_c}{12 \pi^2}
\left[\mu u(\mu^2-\frac{5}{2}m^2)+
\frac{3}{2}m^4\ln{\left(\frac{\mu+u}{m}\right)}
\right]\;,
\label{Omfree}
\eeq
where $N_c$ is the number of colors and $u\equiv \sqrt{\mu^2-m^2}$.

The first-order correction to Eq. (\ref{Omfree}), also known as the 
exchange energy \cite{Free,Baluni,Toimela}, has the form
\bqa
\Omega^{(1)}&=&-2\pi \alpha_s \frac{N_c^2-1}{2} \sumint_{P,Q,K}
\delta(P-Q-K)\times \nonumber \\
&&\frac{{\rm tr}\left[\gamma_{\mu} (P\!\!\!\slash+m_f)
\gamma^{\mu} (Q\!\!\!\slash+m_f)\right]}{(P^2-m_f^2)(Q^2-m_f^2) K^2}\;.
\eqa
We work in the Feynman gauge and have adopted the abbreviation
\beq
\sumint_{P}\equiv T\sum_{p_0} \int \frac{d^3 p}{(2\pi)^3}
\eeq
and 
$\delta(P-Q)=\beta \delta_{p_0,q_0} (2\pi)^3 \delta^3(\vec{p}-\vec{q})$, 
with temperature $T=1/\beta$. Here $p_0=(2n +1) i \pi T+\mu$ is the 
frequency of the quark having four-momentum $P$ and a chemical
potential $\mu$, and the sum is understood to be carried out over the
integers $n$ \cite{Kapusta}.

Using standard quantum field theoretical methods, one obtains 
the complete renormalized exchange energy for a massive quark in the 
$\overline{\hbox{\scriptsize MS}}$ scheme (in the limit $T\rightarrow 0$):
\bqa
\Omega^{(1)}&=&\frac{\alpha_s (N_c^2-1)}{16\pi^3} %
\left[3 \left(m^2 \ln\frac{\mu+u}{m}-\mu u
\right)^2-2 u^4\right.\nonumber\\
&&\left.+m^2\left( 6\ln\frac{\bar{\Lambda}}{m} +4\right) 
\left(\mu u-m^2 \ln\frac{\mu+u}{m}\right)\right]\,.
\label{Om2final}
\eqa
The thermodynamic potential to order $\alpha_s$ for one massive flavor, given 
by the sum of Eqs. (\ref{Om2final}) and (\ref{Omfree}), depends on
the quark chemical potential $\mu$ and on the renormalization subtraction
point $\bar{\Lambda}$ both explicitly and implicitly through the scale
dependence of the strong coupling constant 
$\alpha_s(\bar{\Lambda})$ and the mass
$m(\bar{\Lambda})$.
The scale dependencies of both $\alpha_s$ and $m$, which in the following
we will take to be the mass of the strange quark, are known up to 4-loop
order in the $\overline{\hbox{\scriptsize MS}}$ scheme 
\cite{Vermaseren:1997fq}.
Since we have only determined the free energy to first order in $\alpha_s$, 
we choose
\bqa
\alpha_s(\bar{\Lambda})&=&\frac{4\pi}{\beta_0 L}\left[1-
2 \frac{\beta_1}{\beta_0^2} \frac{\ln{L}}{L}\right] \;,  \nonumber\\
m_s(\bar{\Lambda})&=&\hat{m}_s\left(\frac{\alpha_s}{\pi}\right)^{4/9}
\left[1+0.895062
\frac{\alpha_s}{\pi}\right] \;,
\eqa
where $L=2 \ln{(\bar{\Lambda}/\Lambda_{\overline{\hbox{\scriptsize MS}}})}$,
$\beta_0=11-2 N_f/3$, and $\beta_1=51-19 N_f/3$ and we take $N_f=3$. The 
scale $\Lambda_{\overline{\hbox{\scriptsize MS}}}$ and the 
invariant mass $\hat{m}_s$ are fixed by requiring \cite{PDB}
$\alpha_s\simeq 0.3$ and $m_s\simeq 100~$MeV at $\bar{\Lambda}=2~$GeV;
one obtains $\Lambda_{\overline{\hbox{\scriptsize MS}}}\simeq 380~$MeV
and $\hat{m}_s\simeq 262~$MeV. With these conventions, the only freedom
left is the choice of $\bar{\Lambda}$. 

Note that Eq.(\ref{Om2final}) corresponds to an earlier result 
\cite{strangematter} which was derived in a different renormalization scheme. 
Sticking consistently to the
$\overline{\hbox{\scriptsize MS}}$ scheme
offers however some advantages, e.g. one choice of $\bar{\Lambda}$
allows one to use modern-day restrictions \cite{PDB} on both the 
($\mu$-dependent) values of the strange quark mass and the 
strong coupling $\alpha_s$, whereas the authors of \cite{strangematter}
had to resort to (physically motivated) choices for these. 
Also, our 
calculation is straightforward to
extend to higher orders, since the coefficients for the renormalization
group equations are known up to 4-loop order in the 
$\overline{\hbox{\scriptsize MS}}$ scheme \cite{Vermaseren:1997fq}.

\begin{figure}
\begin{center}
\includegraphics[width=0.9\linewidth]{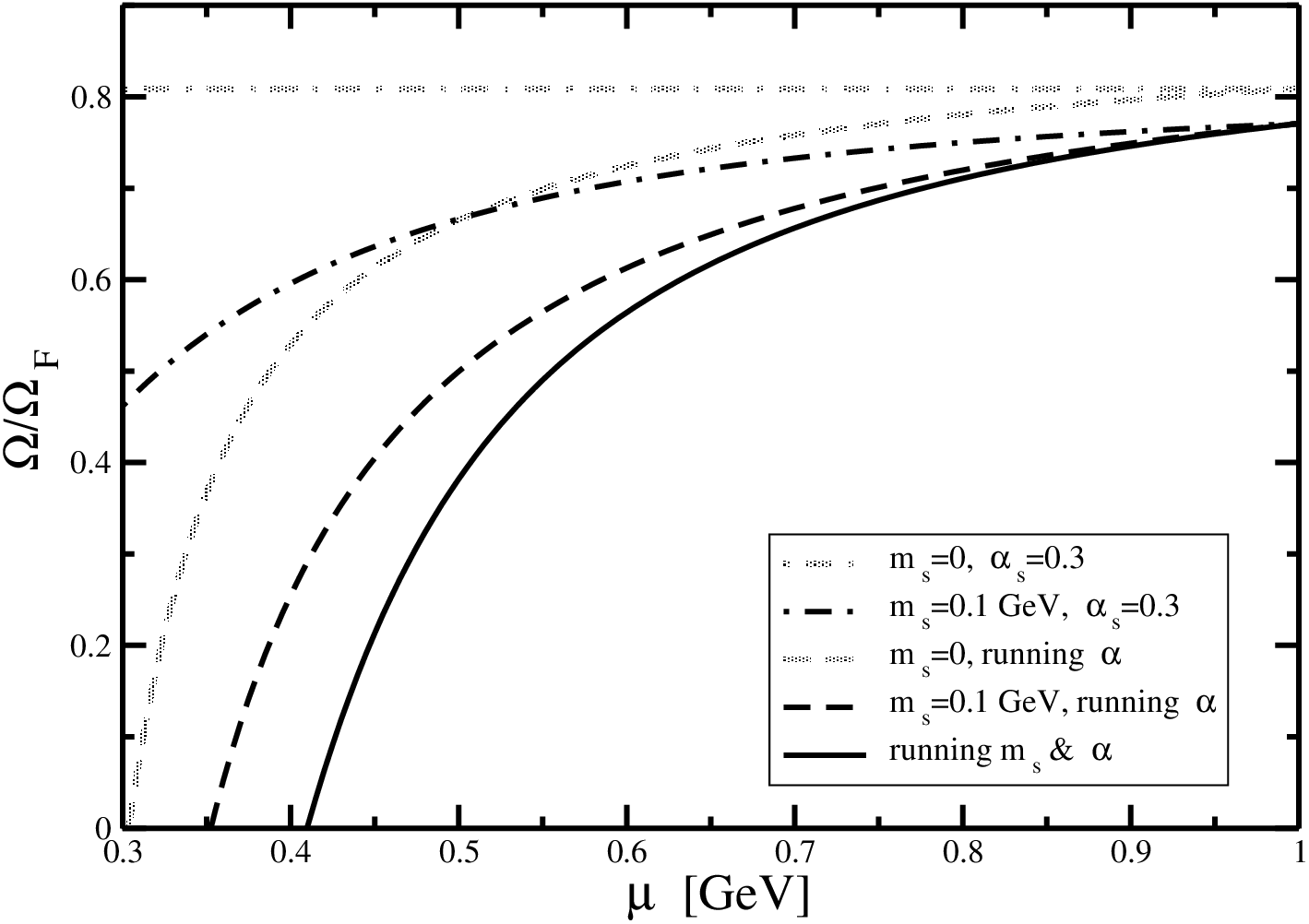}
\setlength{\unitlength}{1cm}
\caption{$\Omega/\Omega_{F}$ at $\bar{\Lambda}=2\mu$ 
for fixed coupling (dash-dotted lines), 
including 2-loop running coupling (dashed lines)
and including running of $m_s$ (full line). The light
gray lines indicate $m_s=0$.}
\label{fig1}
\end{center}
\end{figure}

In Fig. \ref{fig1} we show the thermodynamic  potential for one massive 
flavor (the strange quark), 
normalized by the free (massless) Fermi result, computed in the cases of fixed 
and running coupling and strange quark mass for 
$\bar{\Lambda}=2 \mu$. As one can see, renormalization group improvements 
bring sizeable corrections in the region relevant for the 
structure of compact stars, 
$\mu<1$ GeV. At $\mu=600~$MeV, for which $\alpha_s\sim 0.43$, including a 
fixed mass for the strange quark decreases the thermodynamic potential by 
$\sim 15\%$. The running of the mass (which insures renormalization
scale invariance of $\Omega$ up to order $\alpha_s$) brings plus $\sim 8\%$ of decrease. 
%

To study the effect of the finite strange quark mass on 
the equation of state for 
electrically neutral quark matter with 2 light (massless) 
flavors (up and down quarks) 
and one massive flavor (strange quark), we have to include electrons, with 
chemical potential $\mu_e$. Imposing beta equilibrium, one has
\bqa
&d\rightarrow u+e+\bar{\nu}_e, \quad u+e\rightarrow d+\nu_e,&\nonumber\\
&s\rightarrow u+e+\bar{\nu}_e, \quad u+e\rightarrow s+\nu_e,&\nonumber\\
&s+u\leftrightarrow d+u& .
\eqa
Since neutrinos are lost rather quickly, one may set their chemical
potential to zero \cite{G_b}, so that chemical equilibrium yields
\beq
\mu_d=\mu_s=\mu, \quad \mu_u+\mu_e=\mu \;,
\eeq
with $\mu_u,\mu_d$ and $\mu_s$ the up, down and strange quark chemical
potentials, respectively.
On the other hand, overall charge neutrality requires
\beq
\frac{2}{3} n_u-\frac{1}{3}n_d-\frac{1}{3}n_s-n_e=0 \;,
\eeq
where $n_i$ is the number density of the particle species $i$.
Together, the above equations insure that there is only one independent
chemical potential, which we take to be $\mu$.
Number densities are determined from the thermodynamic potential by
$n_i=-(\partial \Omega/\partial \mu)$
and the total energy density is given by
$\epsilon=\Omega+\sum_i \mu_i n_i$,
where $\Omega=\sum_i \left(\Omega_i^{(0)}+\Omega_i^{(1)}\right)$ and 
again $i$ refers to the particle species. The pressure $P$ is
\beq
P=n_B \frac{\partial \epsilon}{\partial n_B}-\epsilon \;,
\eeq
where $n_B=\frac{1}{3}\left(n_u+n_d+n_s\right)$ is the baryon
number density and the Gibbs potential per particle is given by
$\frac{\partial \epsilon}{\partial n_B}=\left(\mu_u+\mu_d+\mu_s\right)$. 
We restrict the freedom of choice for $\bar{\Lambda}(\mu_u,\mu_d,\mu_s)$ by
requiring that in case of vanishing strange quark mass 
all quark chemical potentials
and number densities become equal so that $P^{(m_s=0)}=-\Omega^{(m_s=0)}$
and, consequently, one has $\mu_e\rightarrow 0$ \footnote{%
This requirement is not automatic since here $\Omega$ is only
renormalization scale invariant up to first order in $\alpha_s$.}.
Furthermore, in order to compare our findings to existing results in the 
literature \cite{Fraga:2001id,tony}, we require 
that in the massless case $\bar{\Lambda}=2\mu$. Consequently,
we choose $\bar{\Lambda}=\frac{2}{3}\left(\mu_u+\mu_d+\mu_s\right)$,
but have tested that our results are not much affected by other choices
obeying the above conditions \footnote{%
Relaxing these conditions one finds that the renormalization scale 
dependence of final results is sizeable as will be shown in \cite{next}; 
however, we want to concentrate on finite mass effects at fixed 
$\bar{\Lambda}$ in this publication.}.

\begin{figure}
\begin{center}
\includegraphics[width=0.9\linewidth]{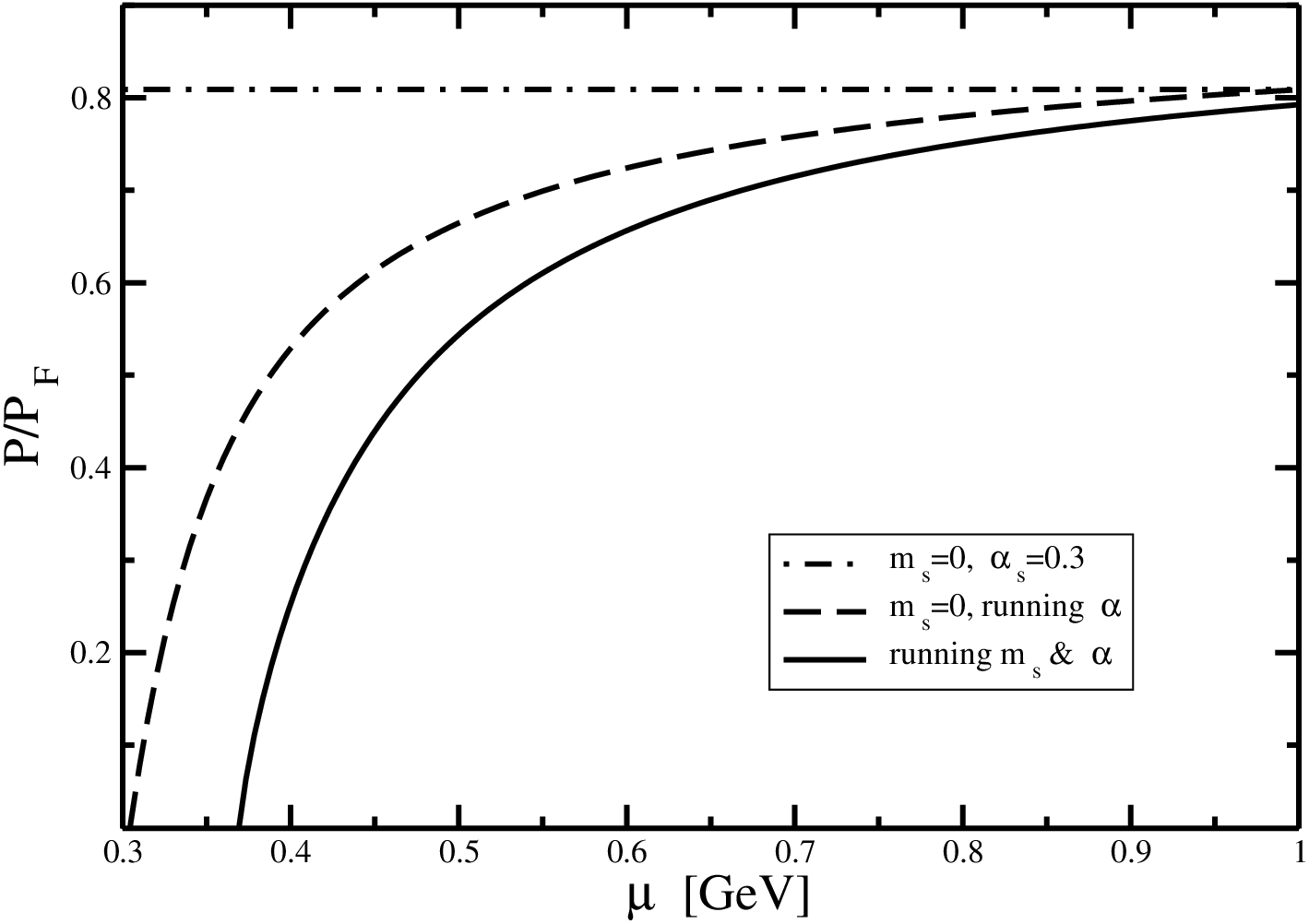}
\end{center}
\vspace*{0.2cm}
\begin{center}
\includegraphics[width=0.9\linewidth]{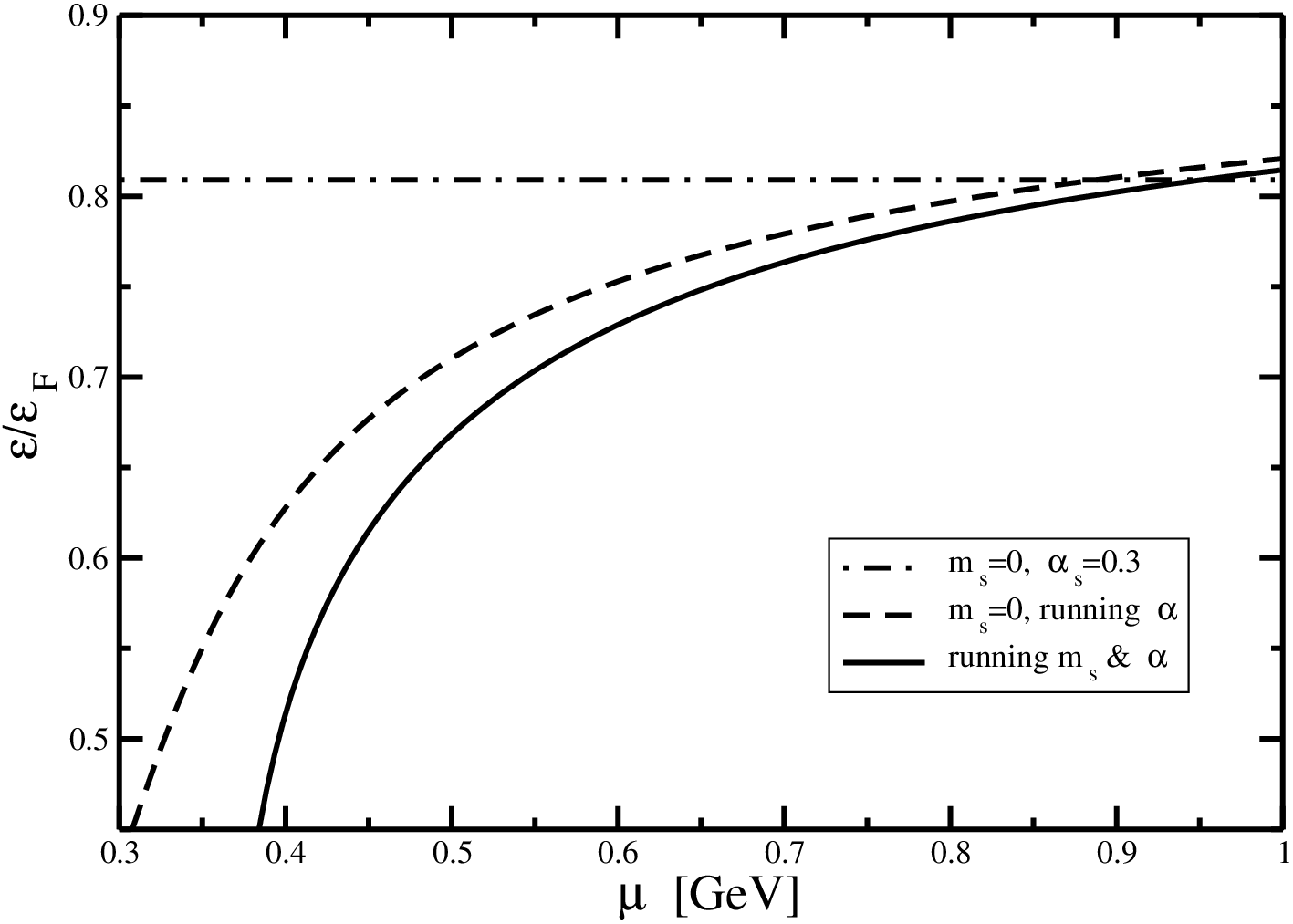}
\setlength{\unitlength}{1cm}
\caption{Pressure and energy density scaled by Fermi values 
for $\bar{\Lambda}=\frac{2}{3}\left(\mu_u+\mu_d+\mu_s\right)$. 
We show results without renormalization group 
improvements (dash-dotted lines), running coupling (dashed
lines), both for $m_s=0$, and results with running mass and coupling 
(full lines).}
\label{fig2}
\end{center}
\end{figure}

The effects of the finite strange quark mass on
the total pressure and energy density for electrically neutral quark matter
(plus electrons) are given in Fig. \ref{fig2}.
There we show results for 3 light flavors and running coupling, corresponding
to the case considered in \cite{Fraga:2001id}, and for 
2 light flavors and one massive flavor, with both
running coupling and strange quark mass (which reaches $m_s\sim 137$ MeV at $\mu=500$ MeV).

As can be seen from this Figure, 
%
there is a
sizable difference between zero and finite strange quark mass pressure
and energy density for the values of the chemical 
potential in the region that 
is relevant for the physics of compact stars, as was already expected from 
the results presented in Fig. \ref{fig1}. As has been noticed by several 
authors \cite{Fraga:2001xc,andre,alford}, the resulting equation of state, 
$\epsilon=\epsilon(P)$, can be approximated by a non-ideal bag model form 
\beq
\epsilon=4 B_{eff}+ a P \;.
\eeq
Here $a\sim 3$ is a dimensionless coefficient while $B_{eff}$ is 
the effective bag constant of the vacuum. Concentrating on the low-density
part of the equation of state, one finds for massless strange quarks 
the parameters $B_{eff}^{1/4} \simeq 117~$MeV and $a\simeq 2.81$ while the
inclusion of the running mass raises these values to 
$B_{eff}^{1/4} \simeq 137~$MeV and $a\simeq 3.17$ (all values having
been obtained by including a running $\alpha_s$ in the equations of state).
Therefore, we expect 
important consequences in the mass-radius relation of quark stars due to 
the inclusion of a finite mass for the strange quark.

\begin{figure}
\begin{center}
\includegraphics[width=0.9\linewidth]{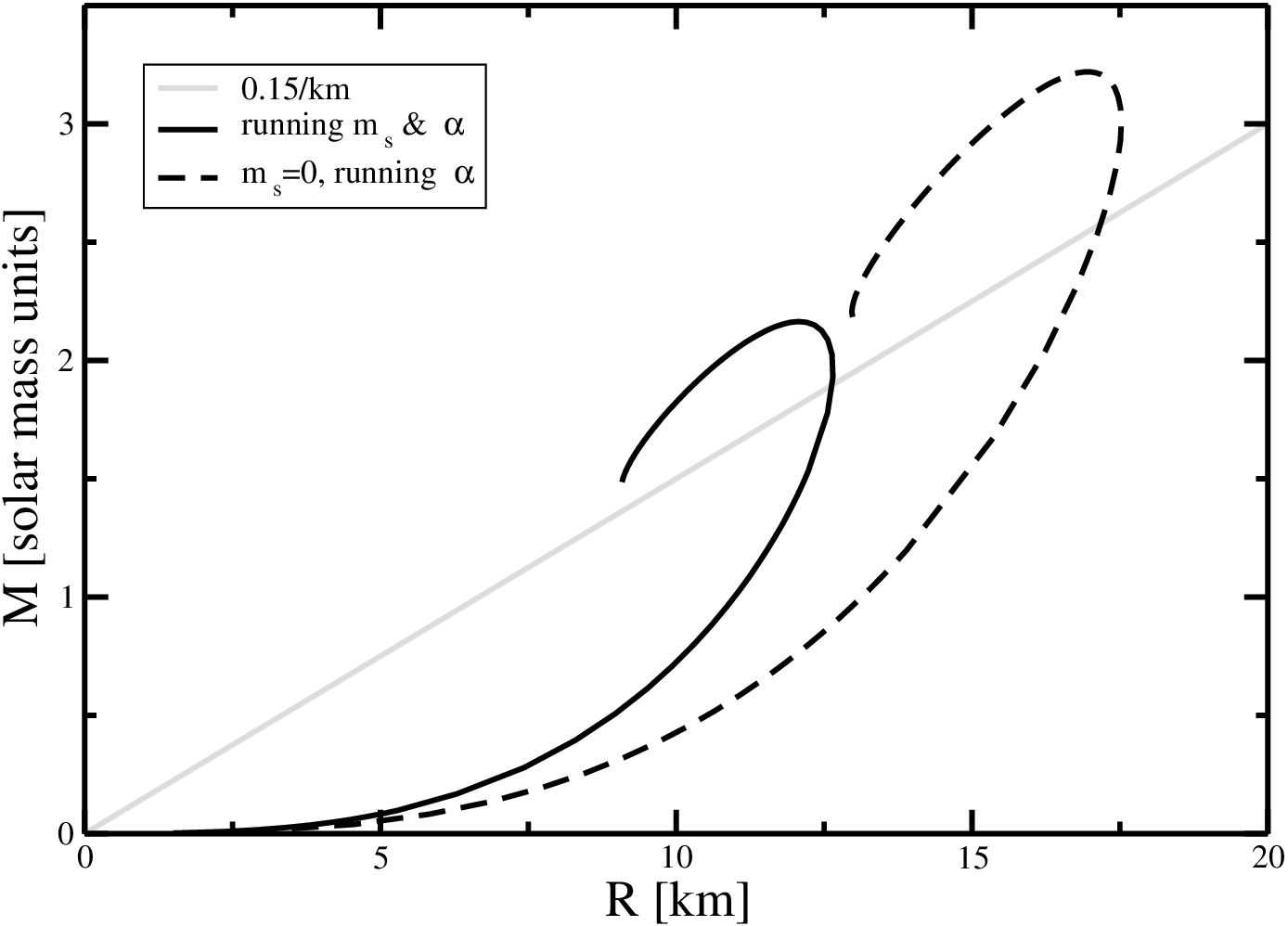}
\setlength{\unitlength}{1cm}
\caption{Mass-radius relation for quark stars based on an equation
of state with running mass and coupling (full line)
and for massless strange quarks 
(dashed line), for running coupling and with $\bar{\Lambda}=\frac{2}{3}\left(%
\mu_u+\mu_d+\mu_s\right)$.}
\label{fig3}
\end{center}
\end{figure}

The structure of a quark star is determined by the solution 
of the TOV equations \cite{G_b}. 
One solves TOV by integrating from some central starting pressure 
up to the surface of the star, where the pressure vanishes. 
The numerical solution of these equations 
in the case of our equation of state for electrically neutral 
quark matter gives the mass-radius relation shown in Fig. \ref{fig3}. 
One can see from the Figure that corrections to the mass and radius 
of quark stars due to a running strange quark mass can be very 
large, $\sim 25\%$. Also, while the most massive star for the
$m_s\!=\!0$ equation of state (with 
$M/M_{\odot}\simeq 3.2$ and radius $\sim 17~$km)
has a central density of $n_B\simeq0.5~{\rm fm}^{-3}$, this number increases
to $n_B\simeq 0.83~{\rm fm}^{-3}$ (at $\mu=470$ MeV) for the heaviest star 
($M/M_{\odot}\simeq 2.16$ at $\sim 12~$km) of the massive equation of state.
One can put constraints on the mass-to-radius ratio by measuring 
the gravitational redshift at the surface of a neutron star. 
Our results for the mass-radius relation are consistent with 
measurements reported by Cottam {\it et al.} \cite{Cottam:2002cu}, 
who obtain a redshift $z=0.35$ for the binary EXO$0748-676$, 
implying a ratio $(M/M_{\odot})/R\sim 0.15/$km, also shown in 
Fig. \ref{fig3}. The inclusion of $\sim \alpha_s^2$ corrections
to the pressure will increase its non-ideality and produce quark
stars which are smaller, denser and less massive 
\cite{Fraga:2001id,Fraga:2001xc}.

We have calculated the thermodynamic potential for 
cold quark matter with two light flavors
and one massive flavor in perturbation theory to first order 
in $\alpha_s$ in the $\overline{\hbox{\scriptsize MS}}$ scheme. 
Results including full corrections $\sim\alpha_s^2$
will appear soon \cite{next} and will certainly modify 
the numbers we present here, since we know that corrections 
are large for this region of chemical potential 
\cite{Fraga:2001id,tony}. Moreover, for a realistic description 
of the structure of quark stars, one has to match the high-density 
equation of state onto a low-density hadronic one. This will 
also shift the values of masses and radii for the stars and, 
in some cases, allow for a new class of compact objects. 
Contributions due to color superconductivity \cite{alford}
as well as chiral condensation \cite{NJL} also affect this 
picture . Finally, if one aims for quantitatively 
tenable results, also the $\sim \alpha_s^3$ contributions to
the pressure should eventually be calculated.
Therefore, our numbers for the maximum mass 
and radius are meant only to be suggestive, and the point that 
we want to stress is the size of the corrections coming from the 
inclusion of a finite strange quark mass.

In this Rapid Communication we present a still rather incomplete picture 
of cold and dense quark matter and the role it plays in the 
core of compact stars. Even so, we believe that our main 
conclusion is qualitatively correct: modifications in the 
pressure due to the inclusion of a nonzero strange quark mass 
are sizable and can dramatically modify the structure of 
quark stars.

\acknowledgments
We would like to thank D. B\"odeker, M. Laine, R.D. Pisarski 
and A. Rebhan for fruitful discussions. 
The authors also thank F. Gelis and E. Iancu for 
their kind hospitality at the Service de Physique Th\'eorique, 
CEA/Saclay, where part of this work has been done. 
The work of E.S.F. is partially supported by CAPES, CNPq, FAPERJ 
and FUJB/UFRJ. P.R. was supported by the DFG-Forschergruppe
Bo 1251/2-2 and KA 1198/4-4.


\begin{thebibliography}{99}

\bibitem{G_b}
N.~K. Glendenning, {\em Compact Stars --- Nuclear Physics, Particle Physics,
and General Relativity} (Springer, New York, 2000);
H.~Heiselberg and M.~Hjorth-Jensen, Phys.\ Rept.\  {\bf 328}, 237 (2000).

\bibitem{Cottam:2002cu}
J.~Cottam, F.~Paerels and M.~Mendez,
Nature {\bf 420}, 51 (2002).

\bibitem{Free}
B.~A.~Freedman and L.~D.~McLerran,
Phys.\ Rev.\  {\bf D16}, 1130 (1977); 
{\it ibid.}, {\bf D16}, 1147 (1977);
{\it ibid.}, {\bf D16}, 1169 (1977);
{\it ibid.},  {\bf D17}, 1109 (1978).

\bibitem{Baluni}
V. Baluni, Phys. Rev. D {\bf 17},  2092  (1978).

\bibitem{Toimela}
T.~Toimela,
Int.\ J.\ Theor.\ Phys.\  {\bf 24}, 901 (1985)
[Erratum-ibid.\  {\bf 26}, 1021 (1987)].

\bibitem{Kapusta}
J.I.~Kapusta,{\em Finite-temperature field theory} (Cambridge University Press,
1989).

\bibitem{wit}
E. Witten, Phys. Rev. D {\bf 30}, 272 (1984).

\bibitem{bag}
C.~Alcock, E.~Farhi and A.~Olinto, 
Astrophys.\ J.\  {\bf 310}, 261 (1986);
P. Haensel, J.~L. Zdunik, and R. Schaeffer, 
Astron. Astrophys. {\bf 160},  121
(1986).

\bibitem{bod}
A.~R. Bodmer, Phys. Rev. D {\bf 4}, 1601 (1971).

\bibitem{Fraga:2001id}
E.~S.~Fraga, R.~D.~Pisarski and J.~Schaffner-Bielich,
Phys.\ Rev.\ D {\bf 63}, 121702 (2001).

\bibitem{Fraga:2001xc}
E.~S.~Fraga, R.~D.~Pisarski and J.~Schaffner-Bielich,
Nucl.\ Phys.\ A {\bf 702}, 217 (2002).

\bibitem{andre}
A.~Peshier, B.~K\"ampfer and G.~Soff,
Phys.\ Rev.\  {\bf C61}, 045203 (2000); 
Phys.\ Rev.\ D {\bf 66}, 094003 (2002).

\bibitem{tony}
J.~P.~Blaizot, E.~Iancu and A.~Rebhan,
Phys.\ Rev.\ D {\bf 63}, 065003 (2001).

\bibitem{andersen}
J.~O.~Andersen and M.~Strickland,
Phys.\ Rev.\ D {\bf 66}, 105001 (2002).

\bibitem{paul}
A.~Rebhan and P.~Romatschke,
Phys.\ Rev.\ D {\bf 68}, 025022 (2003).


\bibitem{alford}
M.~Alford and S.~Reddy,
Phys.\ Rev.\ D {\bf 67}, 074024 (2003); 
M.~Alford, M.~Braby, M.~Paris and S.~Reddy,
nucl-th/0411016.


\bibitem{strangematter}
E.~Farhi and R.~L.~Jaffe, 
Phys.\ Rev.\ D {\bf 30}, 2379 (1984).

\bibitem{next} 
E. S. Fraga and P. Romatschke, in preparation.

\bibitem{NJL}
M.~Buballa and M.~Oertel, 
Nucl.\ Phys.\ A {\bf 703}, 770 (2002);
S.~B.~Ruster {\it et al.},
hep-ph/0503184;
D.~Blaschke {\it et al.},
hep-ph/0503194.


\bibitem{Vermaseren:1997fq}
J.~A.~M.~Vermaseren, S.~A.~Larin and T.~van Ritbergen,
Phys.\ Lett.\ B {\bf 405}, 327 (1997).

\bibitem{PDB}
S.~Eidelman {\it et al.}  [Particle Data Group Collaboration],
Phys.\ Lett.\ B {\bf 592}, 1 (2004).


\end{thebibliography}
\end{document}